\documentclass[review,onefignum,onetabnum]{siamart171218}

\usepackage{lipsum}
\usepackage{amsfonts}
\usepackage{graphicx}
\usepackage{tikz}
\makeatletter
\newcommand{\specificthanks}[1]{\@fnsymbol{#1}}
\makeatother
\usepackage{hyperref}
\usepackage{epstopdf}
\usepackage{algorithmic}
\ifpdf
  \DeclareGraphicsExtensions{.eps,.pdf,.png,.jpg}
\else
  \DeclareGraphicsExtensions{.eps}
\fi
\usepackage{adjustbox}

\newsiamremark{remark}{Remark}
\newsiamremark{hypothesis}{Hypothesis}
\crefname{hypothesis}{Hypothesis}{Hypotheses}
\newsiamthm{claim}{Claim}

\headers{A. Farkane, M. Ghogho, M. Oudani and  M. Boutayeb}{A. Farkane, M. Ghogho, M. Oudani and  M. Boutayeb}

\title{EPINN-NSE: Enhanced Physics-Informed Neural Networks for Solving Navier-Stokes Equations\thanks{
}}

\author{AYOUB FARKANE  \textsuperscript{\specificthanks{3},} \thanks{TICLab, International University of Rabat, Rabat 11100, Morocco.} \href{mailto:ayoub.farkane@uir.ac.ma}{\email{(ayoub.farkane@uir.ac.ma)}}, \hspace*{6mm}\href{mailto:mounir.ghogho@uir.ac.ma}{\email{(mounir.ghogho@uir.ac.ma)}}, \href{mailto:mustapha.oudani@uir.ac.ma}{\email{(mustapha.oudani@uir.ac.ma)}.} 
  \and
 MOUNIR  GHOGHO \footnotemark[2]\and
 MUSTAPHA  OUDANI  \footnotemark[2] \and MOHAMED BOUTAYEB \textsuperscript{\specificthanks{4},} \thanks{CNRS-CRAN-7039, University of Lorraine, France.} \href{mailto:ayoub.farkane@uir.ac.ma}{\email{(mohamed.boutayeb@univ-lorraine.fr).}} \newline \indent \specificthanks{4} INRIA-Nancy-LARSEN, France.}

\usepackage{amsopn}

\makeatletter
\newcommand*{\addFileDependency}[1]{
  \typeout{(#1)}
  \@addtofilelist{#1}
  \IfFileExists{#1}{}{\typeout{No file #1.}}
}
\makeatother

\newcommand*{\myexternaldocument}[1]{%
    \externaldocument{#1}%
    \addFileDependency{#1.tex}%
    \addFileDependency{#1.aux}%
}

\ifpdf
\hypersetup{
  pdftitle={EPINN-NSE: Enhanced Physics-Informed Neural Networks for Solving Navier-Stokes Equations},
  pdfauthor={}
}
\fi

\myexternaldocument{ex_supplement}

\begin{document}

\maketitle

\begin{abstract}
Fluid mechanics is a fundamental field in engineering and science. Solving the Navier-Stokes equation (NSE) is critical for understanding the behavior of fluids. However, the NSE is a complex partial differential equation that is difficult to solve, and classical numerical methods can be computationally expensive. In this paper, we present an innovative approach for solving the NSE using Physics Informed Neural Networks (PINN) and several novel techniques that improve their performance. The first model is based on an assumption that involves approximating the velocity component by employing the derivative of a stream function. This assumption serves to simplify the system and guarantees that the velocity adheres to the divergence-free equation. We also developed a second more flexible model that approximates the solution without any assumptions. The proposed models can effectively solve two-dimensional NSE. Moreover, we successfully applied the second model to solve the three-dimensional NSE.  The results show that the models can efficiently and accurately solve the NSE in three dimensions. These approaches offer several advantages, including high trainability, flexibility, and efficiency.
\end{abstract}



\section{Introduction}
Fluid flow has a wide range of applications in mechanical and chemical engineering, biological systems, and astrophysics.  Essentially, all moving  bodies  are connected to  fluid flow,   such  as vehicles,  airplanes,  trucks, trains, birds, dolphins, and  so  on. 

 Navier-Stokes  equations   are  Partial  Differential Equations (PDEs)   that are used to describe the  flow of fluids.   As with many  PDEs,  the NSE  has no analytical solution, and even in three dimensions, it remains one of the Millennium Prize problems. Classical numerical methods, such as the finite difference method and finite element method (FEM), can produce an approximate solution for the NSE.      
However,  deep learning techniques have recently become popular for solving PDEs. One such method is the physics informed neural network (PINN) \cite{raissi2019physics, raissi2018deep} which uses a deep neural network (DNN) based on optimization problems or residual loss functions to solve a PDE. 
 Other deep learning techniques, such as the deep Galerkin method (DGM)\cite{sirignano2018dgm} have also been proposed in the literature for solving PDEs. The DGM is particularly useful for solving high-dimensional free boundary PDEs such as the High-dimensional Hamilton-Jacobi-Bellman. However, DGM's numerical performance for  other types of PDEs (elliptic, hyperbolic, and partial-integral differential equations, etc.) remains to be investigated.    Different   assumptions are  considered  on the operator  of    PDEs to   guarantee the convergence of this  neural  network. Another approach to approximating PDEs using a variational form including DNNs is the deep Ritz method (DRM) \cite{yu2018deep},  which reformulates the PDE as an optimization problem with a variational formulation.
 The  DRM   is  naturally adaptive, less sensitive to the  problem dimensionality, and works well with the stochastic gradient descent.  Nonetheless, there are a variety of drawbacks to DRM, such as  the problem  of convexity,  and the issue of local minima and saddle points. Moreover,  the DRM has  no consistent conclusion about the convergence rate and the consideration of the necessary boundary condition is more complicated than in previous approaches. Additionally, there are some concerns with the network structure and activation function.

The  work \cite{chan2019machine}  solves a   high dimensional  semi-linear PDE with a condition on diffusion generator of the equation.  To accomplish this, the method used the solution of a backward stochastic differential equation (BSDE) established in \cite{pardoux1990adapted}, and  a time discretization scheme proposed in \cite{bouchard2004discrete}. The approach connected the PDE and stochastic differential equation by using the deep BSDE. However, the numerical results showed that the method did not converge to a standard network of fully connected layers.   Other works are based on the connection between PDEs and stochastic processes through the application of the Feynman–Kac formula \cite{kac1949distributions}.  This paper \cite{beck2021solving} solves a Kolmogorov PDE using deep learning. The Feynman-Kac formula may be extended to the full class of high   linear parabolic equations. 

Chebyshev neural networks are suggested  in \cite{liu2020solving} to  solve   two-dimensional (2D)  linear   PDEs.   The approach involves using a single hidden layer feedforward neural network and approximating the solution with the Kronecker product of two Chebyshev polynomials.  However,  this approach is still far from being  suitable for solving  non-linear PDEs. In some relevant papers \cite{raissi2018hidden, raissi2018numerical}, the nonlinear terms of PDEs were locally linearized in the temporal domain. However, this approach is only suitable for discrete-time domains and may reduce the accuracy of predictions in highly nonlinear systems.  In another approach called MAgNet, proposed in \cite{boussif2022magnet}, a graph neural network was used to forecast a spatially continuous solution of a PDE given a spatial position query. However, MAgNet uses a first-order explicit time-stepping scheme that can be unstable.

It is widely acknowledged that solving nonlinear-dynamic PDEs like the NSE is a challenging task, especially in high dimensions.
The  interesting paper \cite{raissi2019physics}  presented a solution to the 2D NSE using the PINN. In this study, we propose an improved model-based neural network for solving the NSE and compare it to the PINN \cite{raissi2019physics} approach. We also conduct several experiments involving changes in data sizes and hyperparameters. Finally, we apply these enhanced models to solve the three-dimensional (3D) NSE using a test solution.
\section{ Governing Equations}
\subsection{  2D NSE    for Incompressible Flows}
The NSE can be used to model various fluid flows \cite{drazin2006navier}  such as Couette–Poiseuille flows, Beltrami flows, pipe and cylinder flow, and oscillating plates.
Consider the general form of the 2D NSE for an incompressible fluid: 
\begin{align}
\frac{\partial u}{\partial t}+ \beta\left(u \frac{\partial u}{\partial x}+v \frac{\partial u}{\partial y}\right) & =-\frac{\partial p}{\partial 
 x}+ \nu\left(\frac{\partial^2 u}{\partial x^2} + \frac{\partial^2  u }{\partial  y^2}\right) \label{1}\\
\frac{\partial v}{\partial t}+ \beta\left(u \frac{\partial v}{\partial x}+v \frac{\partial v}{\partial y}\right) & =-\frac{\partial p}{\partial 
 y}+ \nu\left(\frac{\partial^2 v}{\partial x^2} + \frac{\partial^2  v }{\partial  y^2}\right), \label{2}
\end{align}
where   $u$ is the x-component of velocity, $v$ is the y-component of velocity, $t$ is time,    $p$ is the flow pressure,  $\nu$ is the fluid viscosity\footnote{
 The viscosity  $\nu$ is  expressed as:
$
\nu=\frac{\rho u L}{R_e}
$
, where
$\rho$  is the fluid density 
, $u$ is  the  fluid velocity
, $R_e$ is the Reynolds number
and $L$  is  the length or diameter of the fluid.} and  $\beta$ is a fixed constant.

The solution of 2D NSE must satisfy the incompressibility equation, which requires the fluid velocity to be a divergence-free function:
\begin{equation}
\frac{\partial u}{\partial x} + \frac{\partial v}{\partial y}=0 \label{3}
\end{equation}
The  previous   equation  describes  the  mass  conservation  of fluid.

When dealing with fluid dynamics, it is necessary to define both the initial and boundary conditions in order to fully specify a boundary value problem. The type of boundary condition used depends on the particular fluid phenomenon being modeled. In some cases, the boundary conditions may be Dirichlet boundary conditions, which specify the values of the solution at the boundaries. In other cases, the boundary conditions may be Neumann boundary conditions, which specify the derivatives of the solution at the borders. It is also possible to use mixed boundary conditions, which combine elements of Dirichlet and Neumann boundary conditions. The selection of the appropriate boundary condition is dependent on the specific problem under consideration.
\subsection{3D NSE}
The problems  \eqref{1}-\eqref{3} can be extended mathematically to three dimensions using the following form:  
\begin{align}
\frac{\partial u}{\partial t}+ \beta\left(u \frac{\partial u}{\partial x}+v \frac{\partial u}{\partial y} + w \frac{\partial u}{\partial z}\right) & =-\frac{\partial p}{\partial 
 x}+ \nu\left(\frac{\partial^2 u}{\partial x^2} + \frac{\partial^2  u }{\partial  y^2}+ \frac{\partial^2  u }{\partial  z^2}\right) \label{4}\\
\frac{\partial v}{\partial t}+ \beta\left(u \frac{\partial v}{\partial x}+v \frac{\partial v}{\partial y} + w \frac{\partial v}{\partial z}\right) & =-\frac{\partial p}{\partial  y}+ \nu\left(\frac{\partial^2 v}{\partial x^2} + \frac{\partial^2  v }{\partial  y^2}\right)+ \frac{\partial^2  v }{\partial  z^2}\label{5}\\
\frac{\partial w}{\partial t}+ \beta\left(u \frac{\partial w}{\partial x}+v \frac{\partial w}{\partial y} + w \frac{\partial w}{\partial z}\right) & =-\frac{\partial p}{\partial  z}+ \nu\left(\frac{\partial^2 w}{\partial x^2} + \frac{\partial^2  w }{\partial  y^2}\right)+ \frac{\partial^2  w }{\partial  z^2}\label{6}
\end{align}
where the velocity components are $u$, $v$, and $w$. The other parameters are defined as previously.  Equation \eqref{6}  introduced the conservation of momentum equation for the $w$ velocity component. As previously stated, the solution can be sought within the divergence-free set:
\begin{equation}
\frac{\partial u}{\partial x} + \frac{\partial v}{\partial y} + \frac{\partial w}{\partial z}=0 \label{7}
\end{equation}
Solving this type of PDE can be challenging, but recent advancements in deep learning techniques have demonstrated potential in addressing these problems. One such approach is the PINN approach, which is elaborated on in the next section.
\section{PINN model}
The PINN is a deep learning algorithm that has been designed to solve PDEs. It combines the flexibility and expressiveness of neural networks with the ability to incorporate physical laws and constraints into the model.
\subsection{Methodology}
The PINN approach consists of two  main  parts. The first  part involves generating a candidate solution, referred to as “$u$”,  at predefined  points  using a multi-layer perception (MLP). The second part consists of calculating the terms associated with the PDE for the candidate solution and constructing a loss function that is optimized using a gradient descent algorithm. We will examine in greater detail how the methods work using the example below. Consider the following one-dimensional PDE:
\begin{align}
\frac{\partial  u  }{\partial  t } +  \frac{\partial^2 u}{ \partial x^2}&= 0, ~~\text{$\forall x$  $\in \Omega$, $\forall t \in [0, T]$}.\\
u=&0  ~~\text{in $\Gamma_{ \Omega}$,  $\forall t \in [0, T]$}\\
u(t=0)&=u_0, ~~\text{$\forall  x \in \Omega$}
\end{align}
In order to apply the PINN to solve the equation shown above, the neural network takes in $x$ and $t$ as inputs and produces a candidate $u$ as the output. The solution is found through minimizing the following loss function:
 \begin{equation}
 \text{MSE}_u= \text{MSE}_0 + \text{MSE}_b + \text{MSE}_f \label{3.4}, 
 \end{equation}
 where: 
\begin{align}
\text{MSE}_0&=\frac{1}{N_0} \sum_{i=1}^{N_0}\left|u\left(0, x_0^i\right)-u_0^i\right|^2\\
\text{MSE}_b &=\frac{1}{N_b} \sum_{i=1}^{N_b}\left(\left|u^i\left(t_b^i,.\right)-u_{b}^i\right|^2\right.\\
\text{MSE}_f&=\frac{1}{N_f} \sum_{i=1}^{N_f}\left|f\left(t_f^i, x_f^i\right)\right|^2
\end{align}
The set of data points that represent the initial conditions of the problem are given by $\left\{x_0^i,u_0^i\right\}_{i=1}^{N_{0}}$, where  $N_0$ is the number of data points; the collocation points on the boundary are represented by the set $\left\{t_b^i\right\}_{i=1}^{N_b}$, where $N_b$ is the number of points; and the set $\{t_f^i, x_f^i\}_{i=1}^{N_f}$ denotes the collocation points on $f(t, x)$.

The mean squared error $\text{MSE}_u$ is a measurement that calculates the average of the squared differences between the predicted and actual values. In the current context, $\text{MSE}_0$ represents the MSE metric specifically associated with the initial conditions of the problem. Similarly, $\text{MSE}_b$ is the MSE metric associated with the boundary conditions, and $\text{MSE}_f$ is a loss function that ensures the solution satisfies the equation within the domain. The function  $f$  is exactly:
 \begin{equation}
     f(t,x)= \frac{\partial u}{\partial t}+ \frac{\partial^2 u}{\partial x^2}
 \end{equation} 
The combination of these three loss functions plays a crucial role in obtaining an accurate solution for the problem at hand. This approach can be extended to different types of PDEs, with some exceptions for more complex problems where modifications to the loss function and neural network architectures may be necessary.
\subsection{PINN to  Solve 2D NSE} \label{PINN2d}
A notable study \cite{raissi2018deep} employed the PINN approach to solving the 2D NSE. The authors used an assumption to simplify the equation of incompressibility \eqref{3} by using a stream function $\varphi$ to approximate the $u$ and $v$ components:
\begin{equation}
u=\frac{\partial \varphi}{\partial y}, \quad v=-\frac{\partial \varphi}{\partial x} \label{9}
\end{equation}
The stream function $\varphi$ is calculated as the neural network’s output, such as pressure $p$. 
 \begin{figure}[h!]
     \centering
     \includegraphics[width=12cm, height=7cm]{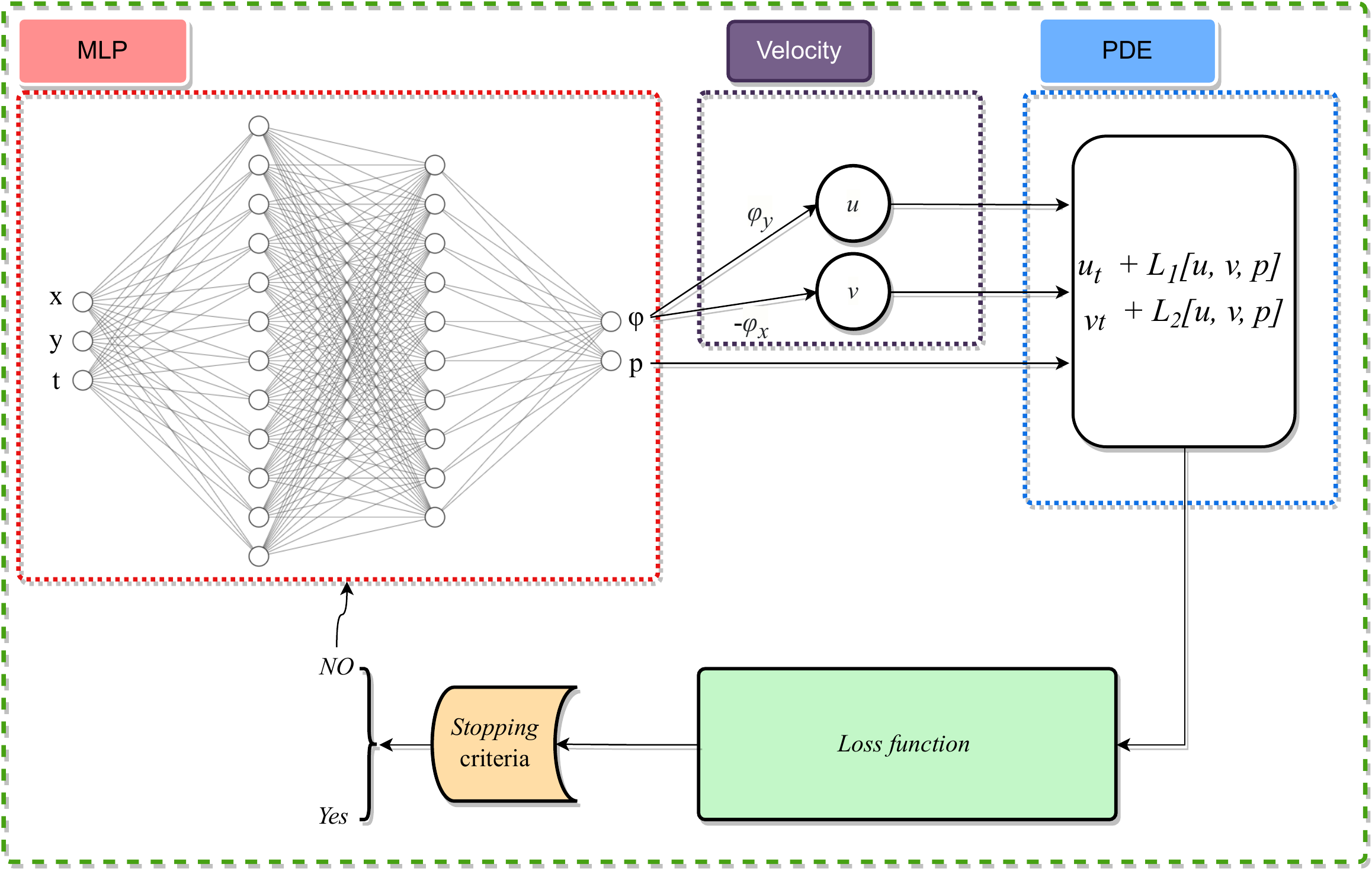}
     \caption{Method with the assumption for finding solutions to the  2D NSE using the PINN. The domain is represented by variables  $x$ and $y$, while $t$ represents time. The stream function is represented by $\varphi$, and its partial derivatives with respect to $x$ and $y$ are represented by $\varphi_x$ and $\varphi_y$, respectively. “Velocity” involves extracting a candidate solution velocity using the assumption. “MLP” is a multi-layer perceptron. “$L_1$, $L_2$” are the derivative operators associated with the equation.}
     \label{f2}
 \end{figure}
Consider the  two functions $f_1$ and $f_2$ such  that:  
 \begin{align}
 f_1(t,x,y)&=\frac{\partial u}{\partial t}+ \beta\left(u \frac{\partial u}{\partial x}+v \frac{\partial u}{\partial y}\right)  +\frac{\partial p}{\partial 
 x}- \nu\left(\frac{\partial^2 u}{\partial x^2} + \frac{\partial^2  u }{\partial  y^2}\right) \label{10}\\
 f_2(t,x,y)&= \frac{\partial v}{\partial t}+ \beta\left(u \frac{\partial v}{\partial x}+v \frac{\partial v}{\partial y}\right) +\frac{\partial p}{\partial 
 y}- \nu\left(\frac{\partial^2 v}{\partial x^2} + \frac{\partial^2  v }{\partial  y^2}\right) \label{11}
 \end{align}
Using the previously defined functions \eqref{10} and \eqref{11}, the model is trained with the following loss functions:  
 \begin{align}
\text{MSE} & =\frac{1}{N} \sum_{i=1}^N\left(\left|u\left(t^i, x^i, y^i\right)-u^i\right|^2+\left|v\left(t^i, x^i, y^i\right)-v^i\right|^2\right) \label{3.12}\\
&\qquad  +\frac{1}{N} \sum_{i=1}^N\left(\left|f_{1}\left(t^i, x^i, y^i\right)\right|^2+\left|f_{2}\left(t^i, x^i, y^i\right)\right|^2\right), \label{3.13}
\end{align}
where the $\left\{t^i, x^i, y^i, u^i, v^i\right\}$ represents accurate data obtained from a PDE simulation or real-world measurements.

The model is trained on data from experiments involving the flow of an incompressible fluid around a circular cylinder. The data is generated using a numerical solver called Nektar++, which discretizes the solution domain into triangular elements and approximates the solution as a linear combination of tenth-order hierarchical, semi-orthogonal Jacobi polynomial expansions.
The method described in this section was used to solve the 2D NSE in this work \cite{raissi2019physics}. While the PINN approach has proven advantageous in addressing this complex problem, current research is focused on refining the model and improving its effectiveness. 
The next section provides a comprehensive overview of the enhanced models, which aim to improve the methodology used to solve the NSE.
\section{EPINN-NSE  Models}
The PINN method is based on deep learning techniques and has shown promise in solving various PDEs. However, further research is necessary to improve its application to more complex problems. There are numerous papers in the literature that focus on improving PINN and its various applications. One of the significant challenges that PINN faces is the issue of training time. While the training duration is not a critical concern for simple cases, it can be a significant issue for complex equations, with training potentially taking days rather than hours. In this context, \cite{sharma2022accelerated} proposed a technique called discretely trained PINNs (DTPINNs) to accelerate the approach by discretizing the PDE and replacing the exact spatial derivatives with high-order numerical discretizations using the finite differences method. Nonetheless, this approach still has a long way to go before it can be applied to problems that involve continuous time.

The PINN approach lacks a theoretical underpinning that ensures the convergence of the model and may be prone to failing to find the appropriate solution.  For instance, the work \cite{krishnapriyan2021characterizing} highlighted some instances where PINN fails to find an accurate solution for certain PDE cases, such as the reaction-diffusion and convection problems. The paper discussed various situations where these networks may fail, such as when the network fails to accurately capture the underlying physics of the system being simulated or when the network is not properly trained. Furthermore, the framework in \cite{krishnapriyan2021characterizing} also provides potential solutions to these issues, such as incorporating more physics information into the network or using different training methods. 
\begin{figure}[h!]
    \centering
    \includegraphics[height=8cm]{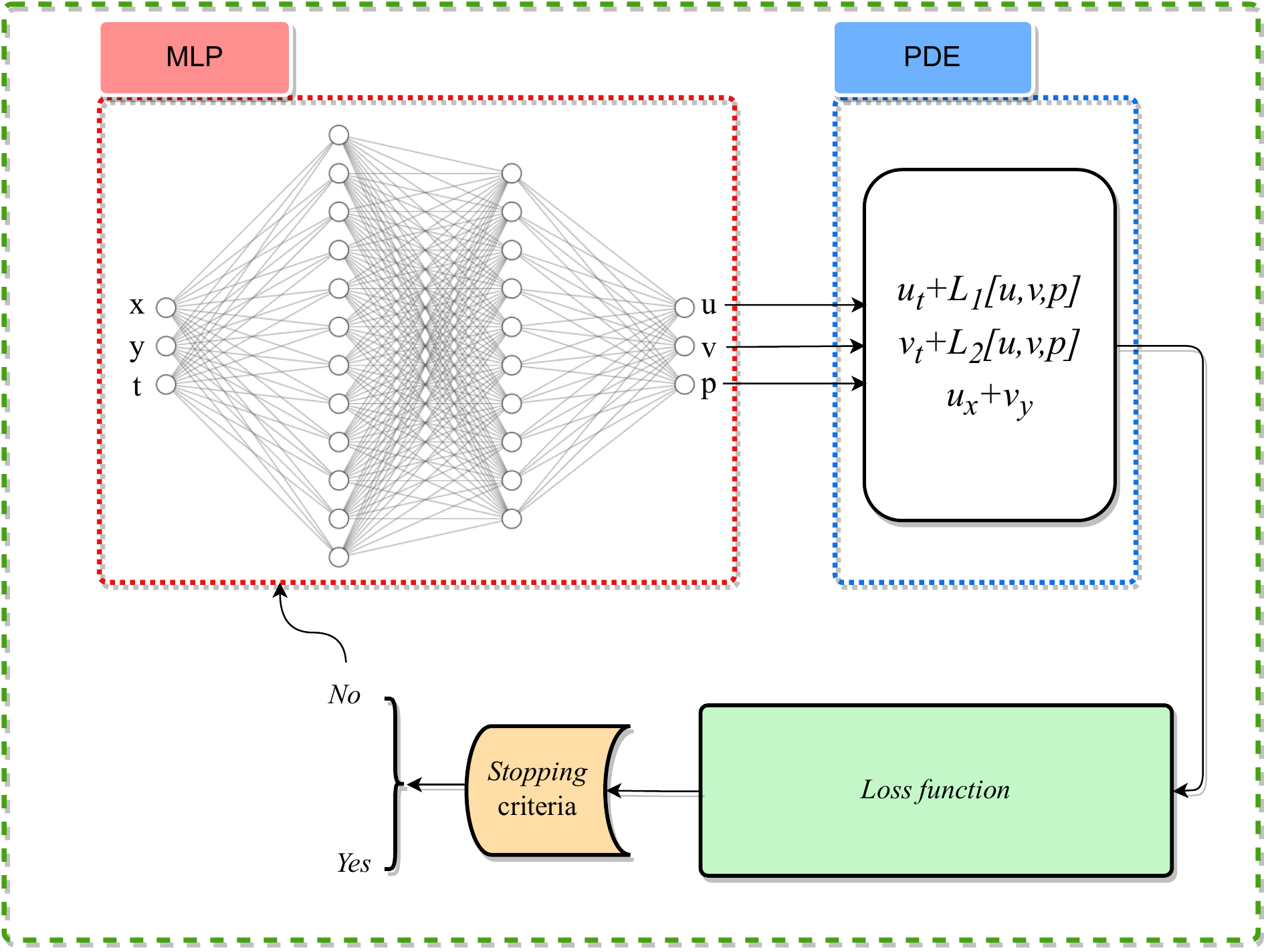}
    \caption{Model of PINN  without using any assumption for solving the 2D NSE.}
    \label{f3}
\end{figure}
Solving the 2D NSE presents additional challenges for PINN. For instance, the model introduced in \cite{raissi2019physics}  required an extended training duration (more than 8 hours).  The  assumption  \eqref{9} is based on the  derivative  relation  between the stream function $\varphi$ and the velocity components respectively $u$, and $v$. The factor of   learning  the  pressure without  training   leads to a higher  probability that the model  produces inaccurate solutions. Another issue related to neural network learning is that the PINN approach uses full batch learning, which may increase the probability of getting stuck on saddle points or local minima. Using full batch learning for training the model may present challenges when working with a large dataset, which can negatively impact the results.

This paper proposes improvements to the previously introduced PINN model for solving PDEs. Specifically, we enhance the model described in Section \ref{PINN2d} by incorporating training on pressure and using advanced deep learning techniques. This involves introducing a pressure loss function to equations \eqref{3.12} and \eqref{3.13}:
\begin{equation}
 \frac{1}{N} \sum_{i=1}^N\left(\left|p\left(t^i, x^i, y^i\right)-p^i\right|^2\right),   \label{p.4} 
\end{equation}
where  the $p\left(t^i, x^i, y^i\right)$ represents  the pressure candidate and $p^i$ pressure  obtained   from  data.
The aim of incorporating training on pressure is  to enhance the pace and accuracy of learning and to improve the results of approximating pressure.  

In addition to those improvements, we develop a novel model that directly approximates the velocity components and pressure without using the assumption  \eqref{9}.  Canceling this assumption based on the derivatives of the stream function can improve the training time and the overall results. The new model employs a neural network that takes space and time variables as inputs and produces solutions for velocity and pressure. The governing equations are computed using automatic differentiation to determine the terms of the loss function that the network will minimize.

 Furthermore, our models incorporate technical improvements in neural network architecture and optimize hyperparameters.  Several studies \cite{khirirat2017mini, hinton2012neural,qian2020impact} have highlighted the advantages of using mini-batch gradient descent. To take advantage of these benefits, we use mini-batch learning in our approach. This enables us to handle large datasets and make use of parallel processing, leading to faster convergence and lower memory requirements. We employed the Adam optimizer \cite{kingma2014adam}, which is  less  sensitive to the initialization of the model parameters. Compared to other optimization algorithms, Adam   is  less likely to get stuck in local minima or saddle points.
 Adam adjusts the learning rate for each parameter based on its past gradients, resulting in faster convergence and improved optimization. To effectively train the models and find the most-effective one, we use the validation loss function to evaluate the models at each epoch and compare the results with previous loss values. Afterward, we save the model that performs best. Moreover, we integrated an early stopping criterion or “patience.” This technique seeks to terminate the model training process when it fails to exhibit significant improvement after a predetermined number of iterations.
\subsection{Solving the 2D NSE}
For this case, we aim to employ   both models to solve  the problem. The first   model  based on  assumption \eqref{9} is  explained in  Figure \ref{f2} with  adding  the 
 techniques introduced previously and  the loss  function associated with pressure  \eqref{p.4}.    In  the  second approach  (Figure \ref{f3}),  the neural network generates a candidate solution $u$, $v$,  and  $p$. This solution is then evaluated using a loss function.
Removing the  assumption  \eqref{9} requires the addition of the loss function linked to the incompressibility equation  \eqref{7}. The neural network is trained to find the solution and  learn the  functions $f_1(t,x,y)$ and $f_2(t,x,y)$ by minimizing  the following  loss function:
\begin{align}
\text{MSE}&= \text{MSE}_e +  \text{MSE}_f + \text{MSE}_d \\
\text{MSE}_e &=\frac{1}{N} \sum_{i=1}^N\left(\left|u\left(t^i, x^i, y^i\right)-u^i\right|^2+\left|v\left(t^i, x^i, y^i\right)-v^i\right|^2\right)  \label{4.1}\\
&\qquad +\frac{1}{N} \sum_{i=1}^N\left(\left|p\left(t^i, x^i, y^i\right)-p^i\right|^2 \right) \nonumber\\
 \text{MSE}_f&=\frac{1}{N} \sum_{i=1}^N\left(\left|f_{1}\left(t^i, x^i, y^i\right)\right|^2+\left|f_{2}\left(t^i, x^i, y^i\right)\right|^2\right) \label{4.2}
\\
\text{MSE}_d&=\frac{1}{N} \sum_{i=1}^N\left(\left|u_{x}\left(t^i, x^i, y^i\right)+v_{y}\left(t^i, x^i, y^i\right)\right|^2\right), \label{4.3}
\end{align}
where $N$  is the number of collocation  points denoted  as  $\left\{t^i,~ x^i,~ y^i\right\}$ and   $u(t^i,x^i,y^i)$, $v(t^i, x^i, y^i)$ and  $p(t^i, x^i, y^i)$    are  the candidate solutions being evaluated.  $u^i$, $v^i$, and $p^i$ are  the real solutions obtained  from a PDE  solver  or real-world observations. 

The loss function specified in equation \eqref{4.1} is employed to measure the difference between the exact and approximate solutions. Additionally, the loss function defined in equation \eqref{4.2} ensures that the solution adheres to the equation, while the loss function outlined in equation \eqref{4.3} represents the divergence-free equation.
\subsection{Solving the 3D NSE}
The 3D NSE is one of the most difficult PDEs to solve.  Only a few studies have used deep learning techniques to tackle this equation, particularly in a 3D problem.  Here we  cite some  papers \cite{margenberg2022neural, ranade2021discretizationnet} that have been conducted using numerical methods such as the FEM and finite volume method (FVM) integrated with DNNs. Other works \cite{eivazi2022physics, wang2022dense, qiu2022physics} addressed specific cases of the NSE in fluid mechanics problems using PINNs. In  this section,    we  adapted  the $I_2$ model  to  approximate  the solution to the 3D problem.
 
Consider the 3D NSE: 
\begin{align}
\frac{\partial u}{\partial t}+ \beta\left(u \frac{\partial u}{\partial x}+v \frac{\partial u}{\partial y} + w \frac{\partial u}{\partial z}\right) +\frac{\partial p}{\partial 
 x}- \nu\left(\frac{\partial^2 u}{\partial x^2} + \frac{\partial^2  u }{\partial  y^2}+ \frac{\partial^2  u }{\partial  z^2}\right)& = h(t,x,y,z)\label{6.11}\\
\frac{\partial v}{\partial t}+ \beta\left(u \frac{\partial v}{\partial x}+v \frac{\partial v}{\partial y} + w \frac{\partial v}{\partial z}\right)+\frac{\partial p}{\partial  y}-\nu\left(\frac{\partial^2 v}{\partial x^2} + \frac{\partial^2  v }{\partial  y^2}+ \frac{\partial^2  v }{\partial  z^2}\right) & = g(t,x,y,z)\label{6.22}\\
\frac{\partial w}{\partial t}+ \beta\left(u \frac{\partial w}{\partial x}+v \frac{\partial w}{\partial y} + w \frac{\partial w}{\partial z}\right)+\frac{\partial p}{\partial  z}-\nu\left(\frac{\partial^2 w}{\partial x^2} + \frac{\partial^2  w }{\partial  y^2}+ \frac{\partial^2  w }{\partial  z^2}\right) &= k(t,x,y,z)\label{6.23}, 
\end{align}
where   $(t, x, y, z)$ $ \in  [0, T]\times\Omega$,  such  that $\Omega \subset \mathbb{R}^3$ (we assume  $\Omega=[0,1]^3$,  $T=20$).  The functions $h$, $g$,  and  $k$ represent the     force components applied  to the system under consideration, While the other variables are as  defined in Section $2$. 
The solution sought satisfies:
\begin{equation}
\frac{\partial u}{\partial x} + \frac{\partial v}{\partial y} + \frac{\partial w}{\partial z}= I(t,x,y,z) \label{6.24}
\end{equation}
The  homogeneous  Dirichlet   boundary   condition  is  considered   for this problem: 
 \begin{equation}
     u (t,x, y,z)=  v(t,x, y,z)=  w(t, x,y,z)=0, ~~  \forall (x, y, z) \in \Gamma_{\Omega}, \forall t \in [0, T],
 \end{equation}
where   $\Gamma_{\Omega}$ is the boundary of domain  $\Omega$.

\begin{figure}[h!]
     \centering
\begin{tikzpicture}[>=latex,font=\fontsize{8}{12}\selectfont]
\draw[rounded corners=5pt,fill=red!18]  (-1.5,-4) rectangle (5.5,1);
\node[draw, rectangle, fill= red!18] at (2,1.3) {MLP};
\node[draw, circle, fill=red!30] (x) at (-2,0) {$x$};
\node[draw, circle, fill=red!30] (y) at (-2,-1) {$y$};
\node[draw, circle, fill=red!30] (z) at (-2,-2) {$z$};
\node[draw, circle, fill=red!30] (t) at (-2,-3) {$t$};

\node[draw, circle, fill=red!30] (h1) at (0,0) {$h_{11}$};
\node[draw, circle, fill=red!30] (h2) at (0,-1) {$h_{12}$};
\node[draw, circle, fill=red!30] (h3) at (0,-2) {$\vdots$};
\node[draw, circle, fill=red!30] (h4) at (0,-3) {$h_{1n}$};
\node[draw, circle, fill=red!30] (h6) at (2,-0) {$h_{21}$};
\node[draw, circle, fill=red!30] (h9) at (2,-1) {$h_{22}$};
\node[draw, circle, fill=red!30] (h7) at (2,-2) {$\vdots$};
\node[draw, circle, fill=red!30] (h10) at (2,-3) {$h_{2n}$};
\node[draw, circle, fill=red!30] (h11) at (3,-1.5) {$\dots$};
\node[draw, circle, fill=red!30] (h8) at (4,0) {$h_{k1}$};
\node[draw, circle, fill=red!30] (h12) at (4,-1) {$h_{k2}$};
\node[draw, circle, fill=red!30] (h13) at (4,-2) {$h_{k3}$};
\node[draw, circle, fill=red!30] (h14) at (4,-3) {$h_{k4}$};
\draw[->] (x) -- (h1);
\draw[->] (y) -- (h1);
\draw[->] (t) -- (h1);
\draw[->] (z) -- (h1);
\draw[->] (x) -- (h2);
\draw[->] (y) -- (h2);
\draw[->] (t) -- (h2);
\draw[->] (z) -- (h2);
\draw[->] (x) -- (h3);
\draw[->] (y) -- (h3);
\draw[->] (t) -- (h3);
\draw[->] (z) -- (h3);
\draw[->] (x) -- (h4);
\draw[->] (y) -- (h4);
\draw[->] (t) -- (h4);
\draw[->] (z) -- (h4);
\draw[->] (h1) -- (h6);

\draw[->] (h1) -- (h7);
\draw[->] (h1) -- (h9);
\draw[->] (h2) -- (h9);
\draw[->] (h3) -- (h9);
\draw[->] (h4) -- (h9);

\draw[->] (h1) -- (h10);
\draw[->] (h2) -- (h10);
\draw[->] (h3) -- (h10);
\draw[->] (h4) -- (h10);
\draw[->] (h2) -- (h7);
\draw[->] (h2) -- (h6);
\draw[->] (h3) -- (h6);
\draw[->] (h3) -- (h7);
\draw[->] (h4) -- (h7);
\draw[->] (h4) -- (h6);

\node[draw, circle,fill=red!30] (u) at (6,0) {$u$};
\node[draw, circle,fill=red!30] (v) at (6,-1) {$v$};
\node[draw, circle,fill=red!30] (w) at (6,-2) {$w$};
\node[draw, circle,fill=red!30] (p) at (6,-3) {$p$};
\draw[->] (h8) -- (u);
\draw[->] (h12) -- (v);
\draw[->] (h13) -- (w);
\draw[->] (h14) -- (p);

\node[draw, rectangle, fill=cyan!40, minimum width=3cm, minimum height=1cm, rounded corners] (g) at (9,-1.5) {Governing equation};
\draw[->] (u) -- (g);
\draw[->] (v) -- (g);
\draw[->] (w) -- (g);
\draw[->] (p) -- (g);
\node[draw, rectangle, fill= green!30, minimum width=3cm, minimum height=1cm, rounded corners] (h) at (9,-5.5) {Loss function};
\draw[->] (g) -- (h);
\node[draw, rectangle, fill=orange!30, minimum width=3cm, minimum height=1cm, rounded corners] (i) at (5,-5.5) {Stopping criteria};
\draw[->] (h) -- (i);
\draw[->] (i) -- node[midway,above] {No} (2,-5.5)-- (1,-4);
\draw[-|] (i) -- node[midway,left] {Yes} (5,-7.5);
\end{tikzpicture}
\caption{  Approach to solve  3D NSE   using  the $I_2$ model.   $n$ represents  the number  of  neurons  per layer and $k$ represents the number of hidden layers.  }
\label{f4}
 \end{figure}
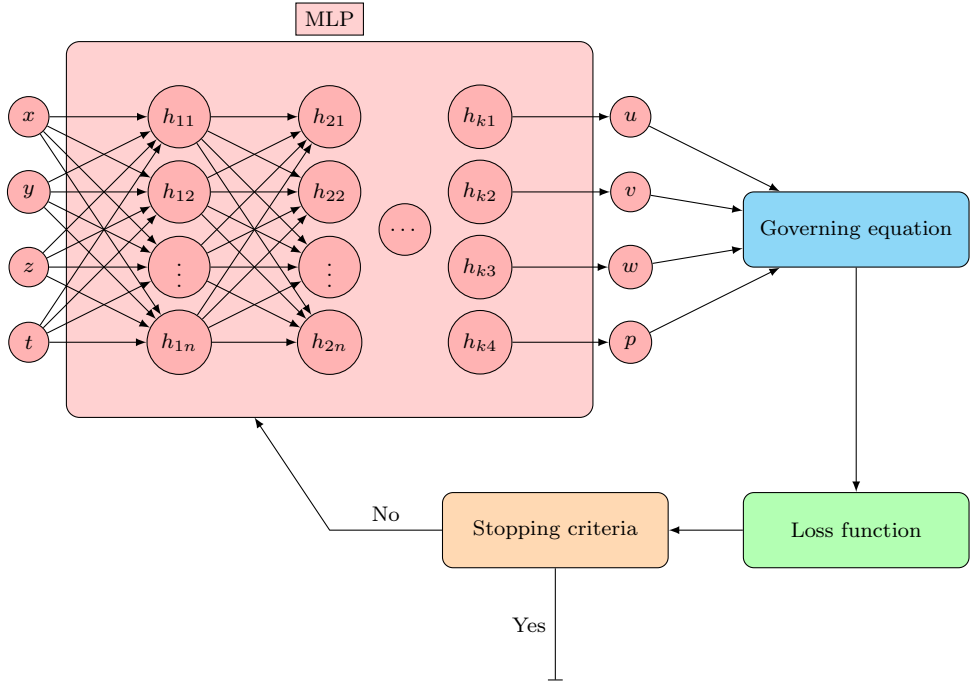

For 3D NSE, as illustrated in Figure \ref{f4}, the neural network inputs consist of the spatial variables $x, y, z$ and the temporal variable $t$. The neural network outputs comprise the three velocity components, represented by $u$, $v$, and $w$, as well as the pressure $p$. Additionally, a similar approach is used in constructing the loss function as  the 2D case. In  this case,  the loss functions are represented as:
\begin{align}
\text{MSE} &= \text{MSE}_{e} + \text{MSE}_{f}+ \text{MSE}_{d}\\
\text{MSE}_{e}&= \frac{1}{N} \sum_{i=1}^N\left(\left|u\left(t^i, x^i, y^i, z^i \right)-u^i\right|^2+\left|v\left(t^i, x^i, y^i,  z^i\right)-v^i\right|^2 \right.\\&\qquad + \left.\left|w\left(t^i, x^i, y^i,  z^i\right)-w^i\right|^2\right)
+\frac{1}{N} \sum_{i=1}^N\left(\left|p\left(t^i, x^i, y^i,  z^i\right)-p^i\right|^2\right) \nonumber\\
\text{MSE}_{f}&=\frac{1}{N} \sum_{i=1}^N\left(\left|f_1\left(t^i, x^i, y^i, z^i\right)\right|^2+\left|f_2\left(t^i, x^i, y^i, z^i\right)\right|^2+\left|f_3\left(t^i, x^i, y^i, z^i\right)\right|^2\right)\\
\text{MSE}_{d} &= \frac{1}{N} \sum_{i=1}^N\Big(\left|u_x(t^i, x^i, y^i, z^i) + v_y(t^i, x^i, y^i, z^i) + w_z(t^i, x^i, y^i, z^i) \right. \Big. \\
&\qquad \left.\left. - I(t^i, x^i, y^i, z^i)\right|^2\right) \nonumber
\end{align}
The neural network generates outputs identified as $u\left(t^i, x^i, y^i, z^i \right)$, $v\left(t^i, x^i, y^i, z^i\right)$, $w\left(t^i, x^i, y^i, z^i\right)$, and $p\left(t^i, x^i, y^i, z^i\right)$. On the other  hand,   $u^i$, $v^i$, $w^i$, and $p^i$ represent the true solution obtained from the data. Additionally, $f_1$, $f_2$, and $f_3$ are  functions defined    as follows:
\begin{align}
f_{1}(t,x,y,z)&= \frac{\partial u}{\partial t}+ \beta\left(u \frac{\partial u}{\partial x}+v \frac{\partial u}{\partial y} + w \frac{\partial u}{\partial z}\right) +\frac{\partial p}{\partial 
 x}- \nu\left(\frac{\partial^2 u}{\partial x^2} + \frac{\partial^2  u }{\partial  y^2}+ \frac{\partial^2  u }{\partial  z^2}\right)\\&\qquad - h(t,x,y,z) \nonumber\\
 f_{2}(t,x,y,z)&= \frac{\partial v}{\partial t}+ \beta\left(u \frac{\partial v}{\partial x}+v \frac{\partial v}{\partial y} + w \frac{\partial v}{\partial z}\right)+\frac{\partial p}{\partial  y}-\nu\left(\frac{\partial^2 v}{\partial x^2} + \frac{\partial^2  v }{\partial  y^2}+ \frac{\partial^2  v }{\partial  z^2}\right) \\&\qquad - g(t,x,y,z)\nonumber\\
 f_{3}(t,x,y,z)&=\frac{\partial w}{\partial t}+ \beta\left(u \frac{\partial w}{\partial x}+v \frac{\partial w}{\partial y} + w \frac{\partial w}{\partial z}\right)+\frac{\partial p}{\partial  z}-\nu\left(\frac{\partial^2 w}{\partial x^2} + \frac{\partial^2  w }{\partial  y^2}+ \frac{\partial^2  w }{\partial  z^2}\right) \\&\qquad - k(t,x,y,z) \nonumber   
\end{align}

 The dataset used to train and evaluate the model’s performance was generated using a test solution, which is given by: 
 \begin{align}
 u&:=(t, x, y, z) \mapsto \mathrm{e}^{-t} \sin (\pi x) \sin (\pi y) \sin (\pi z) \label{6.1}\\
 v&:=(t, x, y, z) \mapsto\mathrm{e}^{-t}\left(x^2-x\right)\left(y^2-y\right)\left(z^2-z\right) \label{6.2}\\
 w&:=(t, x, y, z) \mapsto\mathrm{e}^{-t} \sin (\pi x) \sin (\pi y)\left(z^2-z\right)\label{6.3}\\
 p&:=(t, x, y, z) \mapsto \mathrm{e}^{-t} x y z \label{6.4}
 \end{align}

\section{Experimental Results and Analysis}
In this section, we assess the performance of the improved models in solving the NSE in both 2D and 3D spaces.
\subsection{2D Problem}
\subsubsection{Dataset}
In this study, the data used to train and validate the model is  obtained  from a computational fluid dynamics simulation for incompressible fluid flow around a circular cylinder. The simulation was performed using the Nektar++ \cite{cantwell2015nektar++} PDE solver, as described in the paper \cite{raissi2019physics}. The simulation was carried out in a rectangular domain  $\Omega=[-15,25] \times [-8,8]$, with a constant velocity applied to the left boundary, zero pressure assumed at the right boundary, and periodic conditions applied to the top and bottom boundaries. The simulation is based on the Reynolds number, which is calculated as the ratio of the free stream velocity, cylinder diameter, and kinematic viscosity ($R=$$\frac{U_ {\infty} L}{\nu}$, assumed $U_\infty = 1$, $L=1$, $\nu =0.01$ ). There are 5000 points in the data set, with each point representing 200 time moments, and it includes values for the velocity components $u$ and $v$, as well as the pressure $p$. The data is presented as a collection of    $\left\{t^i, x^i, y^i, u^i, v^i, p^i\right\}$ values.
\subsubsection{Implementations Details}  We conducted several experiments\footnote{The implementations were  executed using the Tesla T4 GPU.} to evaluate the model's performance. These experiments involved analyzing various factors that influence the model’s ability by examining their loss functions, validation loss function, relative errors of predicted solutions, and the training duration required to learn and predict solutions. The objective of these experiments is to extract subsets of data from the global dataset. 
\begin{table}[h!]
\centering
\begin{tabular}{lcc}
\hline
                  & $I_0$                                     &$ I_2$                                \\ \hline
Software library  & TF1.x                                    & Pytorch                             \\ \hline
Dataset          & 1000000                                  & 8334                                \\ \hline
Training data     & 5000                                     & 5000                                \\ \hline
Validation data   & -                                        & 1667                                \\ \hline
Iterations-Epochs & 200000                                   & \textbf{2000}                       \\ \hline
Batch learning    & full-batch                               & \textbf{mini-batch}                 \\ \hline
Batch size        & 5000                                     & 64                                  \\ \hline
Training loss     & 2.825$\times10^{-1}$                                 & \textbf{1.2$\times10^{-2}$}                     \\ \hline
Validation loss   & -                                        & \textbf{6.03e-5}                    \\ \hline
Training duration & \multicolumn{1}{l}{\textgreater{}8h} & \multicolumn{1}{l}{\textbf{1h40min}} \\ \hline
GPU               & Yes                                      & Yes                                 \\ \hline
\end{tabular}
\caption{The initial  training  comparison of $I_0$ and $I_2$.}  \label{t1}
\end{table}

We consider the following notation:
\begin{itemize}
\item $I_0$: represents the initial implementation  of PINN  established in  \cite{raissi2019physics}.
    \item $I_1$: represents the implementation associated  with the   developed model based on assumption \eqref{9}.
    \item $I_2$: represents the implementation  of the model without using  any assumption.
\end{itemize}

Initially, we compared the training performance of one of our proposed models with the approach described in \cite{raissi2019physics}. 

Table \ref{t1} presents a comparison between the initial  model $I_0$,  introduced in paper \cite{raissi2019physics}, and one of the models suggested in this paper $I_2$.  The model  $I_2$ is trained  using Pytorch, while  the  other model uses the TensorFlow $1.x$.  Both models are trained with  the same GPU. 
For   training,  $I_0$   randomly  selected $5000$   points  from the global dataset.  To compare with    $I_2$, we  choose $8334$ points  from the global  data  and   take  $60\% $ ($\simeq$ $5000$) of this data  for training and   $20\%$ ($\simeq$ $1667$) for validation. 
As Table \ref{t1}  shows, $I_2$ outperforms $I_0$ using just 2000 epochs, with lower training loss and shorter training time; $I_0$ required 200,000 iterations.

Using a full batch for training model $I_0$ can create challenges when trying to conduct additional experiments. While more data can be added to allow for longer training times, this also raises the risk of the optimization problem getting trapped at a saddle point or local minimum. To address this, we decided to implement the same model using mini-batch learning, validation loss, and early stopping criteria to alleviate these concerns. Additionally, we further improved the model, which we refer to as model $I_1$, by including training on pressure.
\begin{table}[]
\centering
\begin{tabular}{cc}
\hline
\multicolumn{1}{l}{\textbf{Parameters}} & \multicolumn{1}{l}{\textbf{Details}} \\ \hline
Global data size                       & 5000*200                             \\ \hline
Hidden layer                           & 8                                    \\ \hline
Neurons per layer                      & 20                                   \\ \hline
Activation function                    & Tanh                                 \\ \hline
Optimizer                              & Adam                                 \\ \hline
Initial learning rate                  & 0.001                                \\ \hline 
Early  stopping criteria               & 100                                   \\
\hline
Training data size                     & 60\%                                   \\ \hline
Validation data size                   & 20\% \\ \hline
Test data size                         & 20\%                                   \\ \hline
GPU                                    & Yes                                  \\ \hline
\end{tabular}
\caption{Parameters  shared  in both  models $I_1$ and $I_2$.} \label{t2}
\end{table}

Table \ref{t2} illustrates the common parameters between the two models $I_1$ and $I_2$. These parameters include the neural network’s hyperparameters and the optimization parameters. It should be noted that both models were built from scratch using the PyTorch software library. 
\begin{table}[]
\centering
\begin{tabular}{l|llllll}
\hline
Experiment                                                        & \multicolumn{2}{c|}{$E_{1}$}                                   & \multicolumn{2}{c|}{$E_{2}$}                                   & \multicolumn{2}{c}{$E_{3}$}                                         \\ \hline
Model                                                            & \multicolumn{1}{c|}{$I_{1}$}    & \multicolumn{1}{c|}{$I_{2}$}    & \multicolumn{1}{c|}{$I_{1}$}    & \multicolumn{1}{c|}{$I_{2}$}    & \multicolumn{1}{c|}{$I_{1}$}             & \multicolumn{1}{c}{$I_{2}$} \\ \hline
\begin{tabular}[c]{@{}l@{}}Data \\ size\end{tabular}              & \multicolumn{2}{c|}{9000}                                   & \multicolumn{2}{c|}{10000}                                  & \multicolumn{2}{c}{20000}                                        \\ \hline
\begin{tabular}[c]{@{}l@{}}Early stopping\\ criteria\end{tabular} & \multicolumn{6}{c}{100}                                                                                                                                                                      \\ \hline
\begin{tabular}[c]{@{}l@{}}Batch \\ size\end{tabular}             & \multicolumn{6}{c}{256}                                                                                                                                                                      \\ \hline
\begin{tabular}[c]{@{}l@{}}Training \\ Loss\end{tabular}          & \multicolumn{1}{l|}{1.62e-3} & \multicolumn{1}{l|}{1.65e-3} & \multicolumn{1}{l|}{1.06e-3} & \multicolumn{1}{l|}{1.53e-3} & \multicolumn{1}{l|}{6.98e-4} & 1.38e-3                   \\ \hline
\begin{tabular}[c]{@{}l@{}}Validation \\ loss\end{tabular}        & \multicolumn{1}{l|}{1.66e-3} & \multicolumn{1}{l|}{1.86e-3} & \multicolumn{1}{l|}{1.33e-3} & \multicolumn{1}{l|}{1.78e-3} & \multicolumn{1}{l|}{6.81e-4} & 1.18e-3                   \\ \hline
$RE_{u}$                                                             & \multicolumn{1}{l|}{2.19e-2} & \multicolumn{1}{l|}{2.26e-2} & \multicolumn{1}{l|}{2.03e-2} & \multicolumn{1}{l|}{2.44e-2} & \multicolumn{1}{l|}{1.44e-2} & 1.84e-2                   \\ \hline
$RE_{v}$                                                             & \multicolumn{1}{l|}{7.07e-2} & \multicolumn{1}{l|}{6.14e-2} & \multicolumn{1}{l|}{5.59e-2} & \multicolumn{1}{l|}{6.45e-2} & \multicolumn{1}{l|}{3.99e-2} & 5.12e-2                   \\ \hline
$RE_{p}$                                                             & \multicolumn{1}{l|}{7.30e-2} & \multicolumn{1}{l|}{8.15e-2} & \multicolumn{1}{l|}{7.30e-2} & \multicolumn{1}{l|}{8.26e-2} & \multicolumn{1}{l|}{4.81e-2} & 5.87e-2                   \\ \hline
\begin{tabular}[c]{@{}l@{}}Training\\ duration\end{tabular}       & \multicolumn{1}{l|}{1h15min} & \multicolumn{1}{l|}{28min}   & \multicolumn{1}{l|}{1h45min} & \multicolumn{1}{l|}{32min}   & \multicolumn{1}{l|}{2h16min}          & 46min                     \\ \hline
\end{tabular}

\caption{ Numerical results of experiments $E_1, E_2$ and  $E_3$ for  solving  the 2D NSE using  models $I_1$and $I_2$. Both models 
are  executed with the same early stopping  criteria and  batch size.  }
\label{t3}
\end{table}

In order to evaluate the precision of the  models, we calculate the Relative Error (RE) between the predicted solution and the actual solution. RE is calculated using the following formula: 
\begin{equation}
RE_u=\frac{||predicted~~ value (u) - true ~~value (u)||}{  ||true~~ value(u)||} 
\end{equation}

To conduct additional experiments with larger datasets, we increased the batch size for both implementations, as shown in Table \ref{t4}, to prevent excessive training times. 
\begin{table}[]
\centering
\begin{tabular}{l|llllll}
\hline
Experiment                                                        & \multicolumn{2}{c|}{$E_{4}$}                                   & \multicolumn{2}{c|}{$E_{5}$}                                   & \multicolumn{2}{c}{$E_{6}$}                                \\ \hline
Model                                                             & \multicolumn{1}{c|}{$I_{1}$}    & \multicolumn{1}{c|}{$I_{2}$}    & \multicolumn{1}{c|}{$I_{1}$}    & \multicolumn{1}{c|}{$I_{2}$}    & \multicolumn{1}{c|}{$I_{1}$}    & \multicolumn{1}{c}{$I_{2}$} \\ \hline
\begin{tabular}[c]{@{}l@{}}Data \\ size\end{tabular}              & \multicolumn{2}{c|}{30000}                                  & \multicolumn{2}{c|}{40000}                                  & \multicolumn{2}{c}{60000}                               \\ \hline
\begin{tabular}[c]{@{}l@{}}Early stopping\\ criteria\end{tabular} & \multicolumn{6}{c}{100}                                                                                                                                                             \\ \hline
\begin{tabular}[c]{@{}l@{}}Batch \\ size\end{tabular}             & \multicolumn{6}{c}{500}                                                                                                                                                             \\ \hline
\begin{tabular}[c]{@{}l@{}}Training \\ Loss\end{tabular}          & \multicolumn{1}{l|}{7.26e-4} & \multicolumn{1}{l|}{7.11e-4} & \multicolumn{1}{l|}{4.02e-4} & \multicolumn{1}{l|}{7.33e-4} & \multicolumn{1}{l|}{6.62e-4} & 7.00e-4                   \\ \hline
\begin{tabular}[c]{@{}l@{}}Validation \\ loss\end{tabular}        & \multicolumn{1}{l|}{6.85e-4} & \multicolumn{1}{l|}{6.82e-4} & \multicolumn{1}{l|}{3.69e-4} & \multicolumn{1}{l|}{6.41e-4} & \multicolumn{1}{l|}{5.44e-4} & 6.09e-4                  \\ \hline
$RE_{u}$                                                             & \multicolumn{1}{l|}{1.37e-2} & \multicolumn{1}{l|}{1.30e-2} & \multicolumn{1}{l|}{9.20e-3} & \multicolumn{1}{l|}{1.30e-2} & \multicolumn{1}{l|}{1.19e-2} & 1.30e-2                  \\ \hline
$RE_{v}$                                                             & \multicolumn{1}{l|}{3.93e-2} & \multicolumn{1}{l|}{3.82e-2} & \multicolumn{1}{l|}{2.85e-2} & \multicolumn{1}{l|}{3.61e-2} & \multicolumn{1}{l|}{3.66e-2} & 3.48e-2                  \\ \hline
$RE_{p}$                                                             & \multicolumn{1}{l|}{4.53e-2} & \multicolumn{1}{l|}{4.30e-2} & \multicolumn{1}{l|}{4.29e-2} & \multicolumn{1}{l|}{3.99e-2} & \multicolumn{1}{l|}{4.23e-2} & 4.10e-2                  \\ \hline
\begin{tabular}[c]{@{}l@{}}Training\\ duration\end{tabular}       & \multicolumn{1}{l|}{2h30min} & \multicolumn{1}{l|}{1h12min} & \multicolumn{1}{l|}{3h30min} & \multicolumn{1}{l|}{1h40min} & \multicolumn{1}{l|}{2h55min} & \small 1h30min                  \\ \hline
\end{tabular}
\caption{Numerical outcomes obtained from experiments $E_4, E_5$, and $E_6$ after augmenting the data size and elevating the batch size.}
\label{t4}
\end{table}
We also have further experiments planned in Tables \ref{t5} and \ref{t6} to evaluate the impact of major hyperparameters, such as the number of neurons and layers, on the models. For training and evaluating the models, we used a batch size of 256 and an early stopping criterion of 100 epochs as patience, along with the selection of 20,000 collocation points.
\begin{table}[]
\begin{tabular}{l|llllll}
\hline
Experiment                                                   & \multicolumn{2}{c|}{$E_7$}                                   & \multicolumn{2}{c|}{$E_8$}                                   & \multicolumn{2}{c}{$E_9$}                                \\ \hline
Model                                                        & \multicolumn{1}{c|}{$I_1$}    & \multicolumn{1}{c|}{$I_2$}    & \multicolumn{1}{c|}{$I_1$}    & \multicolumn{1}{c|}{$I_2$}    & \multicolumn{1}{c|}{$I_1$}    & \multicolumn{1}{c}{$I_2$} \\ \hline
Hidden layers                                                & \multicolumn{6}{c}{10}                                                                                                                                                              \\ \hline
\begin{tabular}[c]{@{}l@{}}Neurons per\\ layer\end{tabular}  & \multicolumn{2}{c|}{10}                                     & \multicolumn{2}{c|}{20}                                     & \multicolumn{2}{c}{30}                                  \\ \hline
Training loss                                                & \multicolumn{1}{l|}{3.05e-3} & \multicolumn{1}{l|}{5.21e-3} & \multicolumn{1}{l|}{9.60e-4} & \multicolumn{1}{l|}{9.04e-4} & \multicolumn{1}{l|}{2.96e-4} & 2.67e-4                  \\ \hline
Validation loss                                              & \multicolumn{1}{l|}{2.93e-3} & \multicolumn{1}{l|}{5.00e-3} & \multicolumn{1}{l|}{8.95e-4} & \multicolumn{1}{l|}{9.00e-4} & \multicolumn{1}{l|}{3.19e-4} & 2.62e-4                  \\ \hline
$RE_u$                                                        & \multicolumn{1}{l|}{2.82e-2} & \multicolumn{1}{l|}{4.36e-2} & \multicolumn{1}{l|}{1.65e-2} & \multicolumn{1}{l|}{1.56e-2} & \multicolumn{1}{l|}{8.91e-3} & 7.11e-3                  \\ \hline
$RE_v$                                                        & \multicolumn{1}{l|}{9.72e-2} & \multicolumn{1}{l|}{1.14e-1} & \multicolumn{1}{l|}{4.54e-2} & \multicolumn{1}{l|}{4.37e-2} & \multicolumn{1}{l|}{2.88e-2} & 2.34e-2                  \\ \hline
$RE_p$                                                        & \multicolumn{1}{l|}{1.19e-1} & \multicolumn{1}{l|}{1.41e-1} & \multicolumn{1}{l|}{5.60e-2} & \multicolumn{1}{l|}{4.84e-2} & \multicolumn{1}{l|}{3.08e-2} & 2.78e-2                  \\ \hline
\begin{tabular}[c]{@{}l@{}}Training \\ duration\end{tabular} & \multicolumn{1}{l|}{\small4h40min} & \multicolumn{1}{l|}{\small1h58min} & \multicolumn{1}{l|}{\small4h40min} & \multicolumn{1}{l|}{ \small1h58min} & \multicolumn{1}{l|}{\small2h28min} & \small1h20min                  \\ \hline
\end{tabular}
\caption{Experiments with fixing the  number of hidden layers and changing the number of neurons  per  layer.}
\label{t5}
\end{table}
\begin{table}[h!]
\begin{tabular}{l|llllll}
\hline
Experiment                                                   & \multicolumn{2}{c|}{$E_{10}$}                                   & \multicolumn{2}{c|}{$E_{11}$}                                   & \multicolumn{2}{c}{$E_{12}$}                                \\ \hline
Model                                                        & \multicolumn{1}{c|}{$I_1$}    & \multicolumn{1}{c|}{$I_2$}    & \multicolumn{1}{c|}{$I_1$}    & \multicolumn{1}{c|}{$I_2$}    & \multicolumn{1}{c|}{$I_1$}    & \multicolumn{1}{c}{$I_2$} \\ \hline
Hidden layers                                                & \multicolumn{2}{c|}{12}                                     & \multicolumn{2}{c|}{15}                                     & \multicolumn{2}{c}{18}                                  \\ \hline
\begin{tabular}[c]{@{}l@{}}Neurons per\\ layer\end{tabular}  & \multicolumn{6}{c}{20}                                                                                                                                                              \\ \hline
Training loss                                                & \multicolumn{1}{l|}{1.12e-3} & \multicolumn{1}{l|}{1.49e-3} & \multicolumn{1}{l|}{1.16e-3} & \multicolumn{1}{l|}{8.86e-4} & \multicolumn{1}{l|}{1.27e-3} & 9.40e-4                  \\ \hline
Validation loss                                              & \multicolumn{1}{l|}{1.07e-3} & \multicolumn{1}{l|}{1.38e-3} & \multicolumn{1}{l|}{1.10e-3} & \multicolumn{1}{l|}{8.50e-4} & \multicolumn{1}{l|}{1.11e-3} & 8.69e-4                  \\ \hline
$RE_u$                                                        & \multicolumn{1}{l|}{1.75e-2} & \multicolumn{1}{l|}{2.05e-2} & \multicolumn{1}{l|}{1.86e-2} & \multicolumn{1}{l|}{1.54e-2} & \multicolumn{1}{l|}{1.76e-2} & 1.68e-2                  \\ \hline
$RE_v$                                                        & \multicolumn{1}{l|}{5.11e-2} & \multicolumn{1}{l|}{5.42e-2} & \multicolumn{1}{l|}{5.21e-2} & \multicolumn{1}{l|}{4.25e-2} & \multicolumn{1}{l|}{5.59e-2} & 4.23e-2                  \\ \hline
$RE_p$                                                        & \multicolumn{1}{l|}{6.47e-2} & \multicolumn{1}{l|}{5.85e-2} & \multicolumn{1}{l|}{6.18-e2} & \multicolumn{1}{l|}{5.08e-2} & \multicolumn{1}{l|}{5.90e-2} & 4.96e-2                  \\ \hline
\begin{tabular}[c]{@{}l@{}}Training \\ duration\end{tabular} & \multicolumn{1}{c|}{\small2h20min} & \multicolumn{1}{l|}{1h}      & \multicolumn{1}{c|}{\small3h15min} & \multicolumn{1}{l|}{\small1h43min} & \multicolumn{1}{c|}{\small4h10min} & 2h                       \\ \hline
\end{tabular}
\caption{Outcomes of  fixing the  number  of neurons per layer and changing  the number of layers.}
\label{t6}
\end{table}
Based on the numerical results presented in Tables \ref{t1}, \ref{t3}, \ref{t4} \ref{t5}, and  \ref{t6}, it is evident that the suggested models $I_1$ and $I_2$ outperform other models in solving the 2D NSE. While the $I_1$ model has been considerably enhanced by integrating deep learning techniques, the training process still requires a considerable amount of time. In contrast, the $I_2$ model demonstrates its capability to compete with the $I_1$ model and achieve good results, while requiring a training time that is at most half that of the $I_1$. Furthermore, the results indicate that modifying the architecture and hyperparameters can have a discernible impact on the final results. In fact, as the number of neurons and hidden layers increase, the predicted solution associated with model $I_2$ achieves a higher level of convergence. 

In the next section, we try to solve the non-stationary NSE in a 3D domain using model $I_2$. We selected this approach due to its perceived efficacy in effectively addressing the complexity of the problem at hand.
\subsection{3D Problem}
 \subsubsection{Dataset}
In order to construct a model for the 3D problem,   a dataset comprising variables $\left\{x, y, z, t, u, v, w, p \right\}$ is required.
In this study, a total of 7318 collocation points were generated in the spatial domain using a popular and freely available mesh generator known as Gmsh \cite{geuzaine2009gmsh}.  The time interval was discretized into 200 moments. The solution associated with the points was calculated using the equations  \eqref{6.1}-\eqref{6.4}.

Afterward, the resulting dataset was rearranged into the form  $\left\{x^i, y^i, z^i, t^i, u^i, v^i, w^i, p^i\right\}$ for all $1 \leq i \leq (7318\times200)$, which allowed for more in-depth analysis and interpretation of the model’s results.

\subsubsection{Numerical  Results}  Due to the complexity of the problem, a substantial dataset was utilized for this experiment\footnote{ The Tesla A40 GPU was employed to run the implementations.}.
\begin{figure}[h!]
    \centering
    \includegraphics[height=8cm]{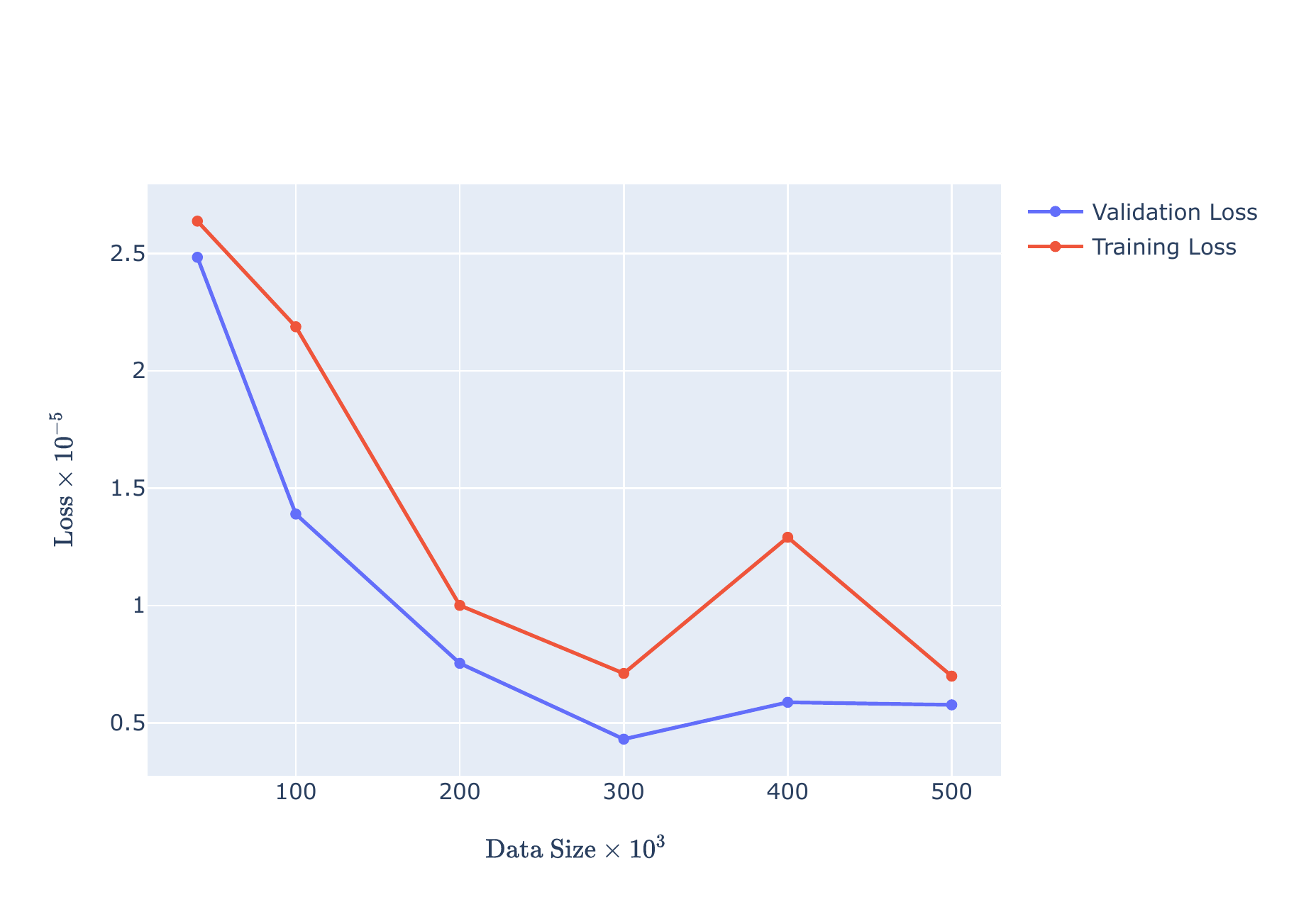}
    \caption{Effects of varying dataset size on training and validation loss.}
    \label{f6}
\end{figure}
\begin{figure}[]
    \centering
    \includegraphics[height=9cm]{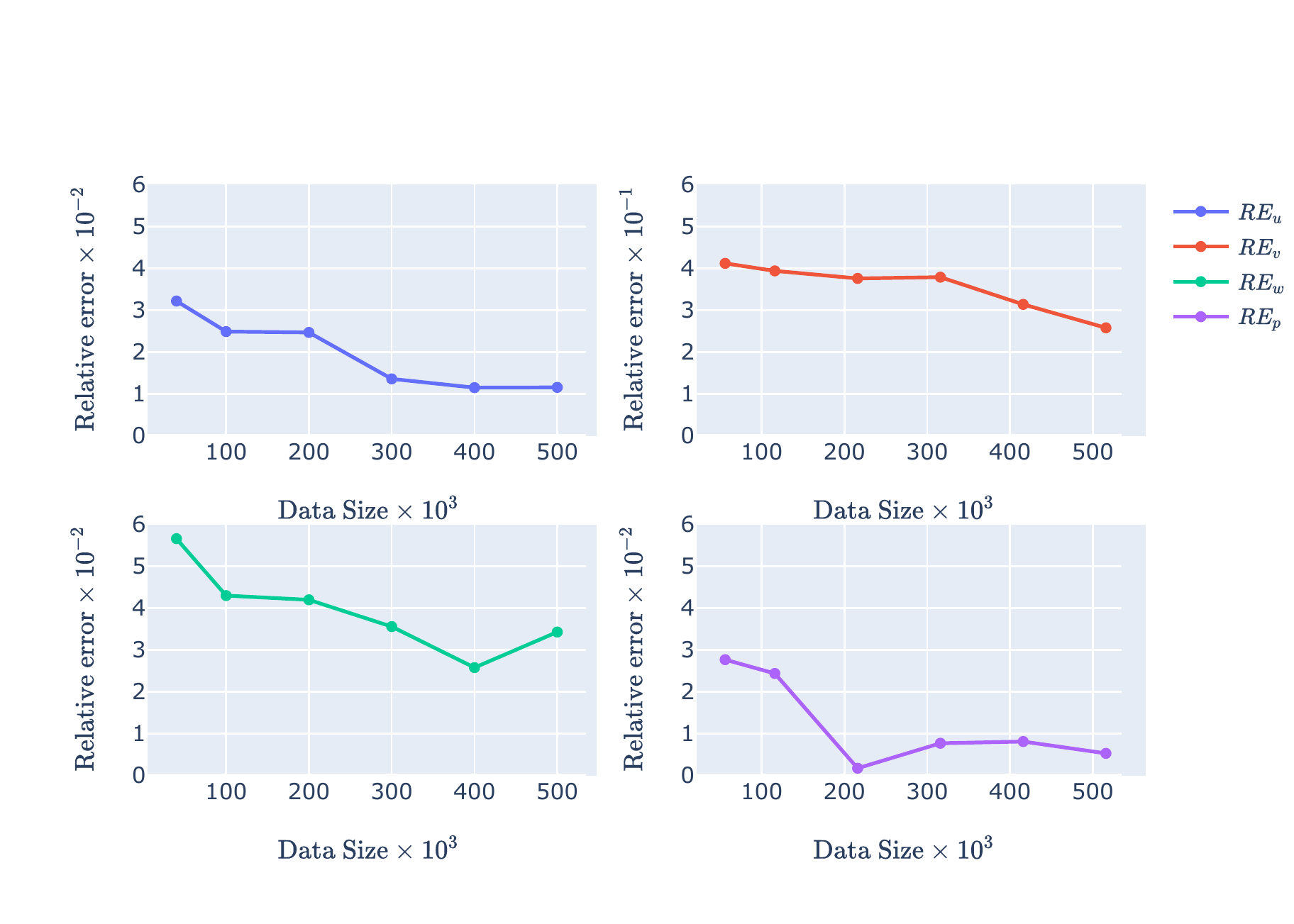}
    \caption{Relative errors of velocity and pressure components in various experiments of 3D NSE solved using the $I_2$ model. }
    \label{f7}
\end{figure}
Figure  \ref{f6}  displays the training and validation losses with respect to the data size.  Based on the conducted experiments, we have successfully run the model using $35\%$ of the entire dataset, which comprises over one million examples. Notably, the model's training can be completed in just 3 hours and 30 minutes using 500,000 examples, demonstrating its excellent efficiency. All experiments were conducted using the same parameters listed in Table  \ref{t2} except the global data size in this case is $7318 \times 200$.

The results presented in Table \ref{t7} were obtained  using the number of epochs as a stopping criterion, with 2000 epochs chosen as the optimal value for training the model. This number was considered large enough to provide sufficient time for the model to optimize and investigate its performance capabilities.
\begin{table}[h!]
\begin{tabular}{l|c|l|l|l|l|c}
\hline
Experiment        & $E_1$    & \multicolumn{1}{c|}{$E_2$}    & \multicolumn{1}{c|}{$E_3$}    & \multicolumn{1}{c|}{$E_4$} & \multicolumn{1}{c|}{$E_5$}   & $E_6$     \\ \hline
Data size         & 40000   & \multicolumn{1}{c|}{100000}  & \multicolumn{1}{c|}{200000}  & 300000                    & \multicolumn{1}{c|}{400000} & 500000   \\ \hline
Loss function     & 2.13e-5 & 7.23e-6                      & 3.46e-6                      & 2.56e-5                   & 1.49e-5                     & 5.37e-6  \\ \hline
Validation loss   & 1.99e-5 & 5.84e-6                      & 3.23e-6                      & 2.75e-6                   & 1.98e-6                     & 2.31e-6  \\ \hline
$RE_u$             & 2.13e-2 & 1.20e-2                      & 1.06e-2                      & 1.13e-2                   & 6.97e-3                     & 5.69e-3  \\ \hline
$RE_v$             & 3.89e-1 & 2.50e-1                      & 4.56e-1                      & 1.94e-1                   & 9.65e-2                     & 8.19e-2  \\ \hline
$RE_w$             & 6.23e-2 & 2.25e-2                      & 2.42e-2                      & 2.48e-2                   & 2.37e-2                     & 1.35e-2  \\ \hline
$RE_p$             & 1.58e-2 & 8.47e-3                      & 4.24e-3                      & 4.98e-3                   & 3.99e-3                     & 2.91e-3  \\ \hline
Training duration & 1h5min  & \multicolumn{1}{c|}{2h36min} & \multicolumn{1}{c|}{5h27min} & \multicolumn{1}{c|}{8h}   & \multicolumn{1}{c|}{11h}    &  \small 13h41min \\ \hline
\end{tabular}
\caption{ The outcomes obtained through the implementation of the $I_2$ model for solving the 3D NSE, wherein the number of epochs was employed as a stopping criterion.   } \label{t7}
\end{table}
By examining  the results shown in Figure \ref{f7}, we can conclude  that the relative errors exhibit a decreasing trend as the data size increases. This observation aligns with the expected behavior of relative errors with increasing data size, as larger datasets enable more precise calculations by providing a greater amount of information. 
\section{Conclusions and  Perspectives}
The NSE represent a highly intricate challenge to solve, given their dynamic and nonlinear nature as PDEs.  Recent research has shown that physics-informed neural networks (PINN) offer a promising solution to this challenge. Our study has demonstrated the effectiveness of improved models based on PINN for solving the NSE in 2D and 3D cases. We have shown that these models offer significant advantages over the original PINN approach, including enhanced computational efficiency, more stable convergence, and reduced relative errors. The approach employed in this paper  involves utilizing a test solution in the  three-dimensional problem for training and evaluating  the efficacy of models aimed at solving complex PDEs. This methodology serves as a viable alternative to conducting real experiments or employing PDE solvers, which may prove more costly.  These findings suggest that further refinement of this method could open up new opportunities for its application to real-world data from experiments in three dimensions. Overall, our work highlights the potential of PINN as a powerful tool for tackling complex physical modeling problems and paves the way for exciting future developments in this area.

 In addition to the promising results presented in this study, there are several exciting perspectives for future research. One possibility is to explore the potential of PINN in solving more complex problems, such as those involving multiphase flows or turbulence. Another avenue for investigation is to extend the methodology developed in this paper to other types of PDEs, including those with non-linear boundary conditions or time-dependent coefficients. Moreover, the use of PINN could be extended to other areas of physics and engineering, such as solid mechanics, electromagnetics, and quantum mechanics, to name a few.

Another perspective to consider is the potential for combining PINN with other machine learning techniques, such as deep learning or reinforcement learning, to further enhance their capabilities. The integration of PINN with other methods may allow for more efficient and accurate solutions to complex physical problems, while also providing insights into the underlying physical phenomena.

Finally, it is essential to recognize the potential impact  of solving PDEs in various fields of science and engineering. The ability to accurately model and predict complex physical phenomena have significant implications for the design of new materials, optimization of manufacturing processes, and the development of new technologies. As such, the continued development and refinement of PDE solvers have the potential to transform how we approach modeling and simulation in various domains, leading to exciting discoveries and advancements.

\bibliographystyle{siamplain}
\bibliography{references}
\end{document}


\maketitle

\section{A detailed example}

Here we include some equations and theorem-like environments to show
how these are labeled in a supplement and can be referenced from the
main text.
Consider the following equation:
\begin{equation}
  \label{eq:suppa}
  a^2 + b^2 = c^2.
\end{equation}
You can also reference equations such as \cref{eq:matrices,eq:bb} 
from the main article in this supplement.

\lipsum[100-101]

\begin{theorem}
  An example theorem.
\end{theorem}

\lipsum[102]
 
\begin{lemma}
  An example lemma.
\end{lemma}

\lipsum[103-105]

Here is an example citation: \cite{KoMa14}.

\section[Proof of Thm]{Proof of \cref{thm:bigthm}}
\label{sec:proof}
\lipsum[106-112]

\section{Additional experimental results}
\Cref{tab:foo} shows additional
supporting evidence. 

\begin{table}[htbp]
{\footnotesize
  \caption{Example table}  \label{tab:foo}
\begin{center}
  \begin{tabular}{|c|c|c|} \hline
   Species & \bf Mean & \bf Std.~Dev. \\ \hline
    1 & 3.4 & 1.2 \\
    2 & 5.4 & 0.6 \\ \hline
  \end{tabular}
\end{center}
}
\end{table}

\bibliographystyle{siamplain}
\bibliography{references}